\documentclass[letterpaper]{jpconf}

\usepackage{graphicx}

\begin{document}

\title{Soft Probes of the Quark-Gluon Plasma with ALICE at LHC}

\author{Renaud Vernet (for the ALICE Collaboration)}

\address{Istituto Nazionale di Fisica Nucleare (INFN), Catania, Italy}

\begin{abstract}
The Large Hadron Collider (LHC) should start its activity of
data taking by the end of summer 2009, and will provide beams of \mbox{p-p} and \mbox{Pb-Pb}
 at colliding energies up to 14~TeV and 5.5~ATeV respectively.
The \mbox{Pb-Pb} heavy-ion program aims at reaching the necessary conditions to create a deconfined state of
partons, the Quark-Gluon Plasma (QGP), whose study is one of the most exciting physics topics
to be explored thanks to the possibilites offered by this
new-generation accelerator. In particular, the "soft" observables related to low and intermediate $p_T$ processes, will shed
light on many fundamental properties of the system, such as thermodynamic parameters,
chemical composition, expansion velocity etc.
The \mbox{p-p} collisions will be of great interest as well, since they will
serve as an essential reference for heavy ions. 
ALICE (A Large Ion Collider Experiment) is the LHC experiment dedicated to the study of the QGP. Its
large acceptance and low magnetic field make it particularly suited
for the study of soft phenomena. After having given an overview of
this detector, I will present the main motivations and prospects for soft physics in both
\mbox{p-p} and \mbox{Pb-Pb} collisions.
\end{abstract}

\section{Introduction}

More than 99\% of the particles produced in heavy-ion collisions have
a momentum lower than $2~{\rm GeV}/c$. Therefore the study of these
``soft'' particles and the physical processes involved in their production
are of major interest for the comprehension of the Quark-Gluon Plasma
(QGP) properties: they give access to the chemical composition of the system, its size, its
temperature, and its dynamics. However, because of the large ``jump''
in energy, the abundancy of hard
processes at LHC will somehow modify our perception of soft physics:
high-$p_T$ mechanisms, which have already left their fingerprints at
RHIC, will certainly influence the medium in many respects.
For this reason, it will be essential to measure all the observables
relevant for heavy-ion data also in small systems such as \mbox{p-p}, first to get a hadronic reference for heavy-ion data
interpretation, and second to understand the underlying pQCD processes 
at these energies.

At LHC, ALICE is the experiment best suited to study soft phenomena
due to its low material budget, its low magnetic field
and its main tracking device (the TPC), which provides a unique capability
to identify many particles down to low momenta $\sim 100~{\rm
MeV}/c$~\cite{Alessandro:2006yt}. 
It is actually the only experiment at LHC almost fully dedicated
to the study of the QGP, though having a substantial physics
program for \mbox{p-p} to achieve.
An overall description of this experiment can
be found in ref.~\cite{Carminati:2004fp}.
The following sections will give an overview of ALICE performance in
the soft sector, showing results obtained for the following topics:
event characterization, (strange-) particle production, flow, femtoscopy and
event-by-event analyses.

\section{\label{sec:eventcharact}Event characterization and particle production}

All the heavy-ion collisions will be sorted as a function
of the impact parameter between the two incident nuclei, used to describe
the event ``centrality''. This is of major importance to derive the
involved number of participant nucleons and binary collisions, which are
critical elements in this field.
%~\cite{Glauber:1959Lecture}, it is used to calculate the number
%of nucleons participating to a collision, 
In ALICE the centrality is measured by combining the information from zero
degree hadronic and electromagnetic calorimeters located near the beam
axis. The distributions of their respective signals
are described in refs.~\cite{Alessandro:2006yt,Ramello:2008zz}.
The correlation of these two signals provide the number of spectators
in the collision, from which one
deduces the number of participants and the impact parameter, with
estimated resolutions of 15 nucleons and $1~{\rm fm}$ respectively~\cite{Alessandro:2006yt}.

Inclusive charged-particle multiplicity and pseudorapidity (${\rm d}N_{ch}/{\rm d}\eta$)
distributions supply additional event information: they
relate to the initial energy density of the medium, and help in
testing particle production models and understanding limiting fragmentation
phenomena (see~\cite{Busza:2007ke}.) 
%~\cite{Nouicer:2002ks}.
In ALICE these measurements are performed using the detectors ${\rm ITS+TPC}$
(central) and FMD (forward region) that allow a coverage of $-3.4<\eta<5.1$~\cite{Riedler:2006jq,Christensen:2007yc}.
The multiplicity and ${\rm d}N_{ch}/{\rm d}\eta$ distributions will be
the very first measurements ALICE will be able to achieve. The first
run should take place in fall 2009 with proton beams, at $\sqrt{s}$
equal to 0.9 and 10~TeV.

The amount of identified particles in ALICE will be sufficiently high to
provide temperature and baryochemical potential event by event. This
will give the possibility to place every event on the nuclear matter
phase diagram ($T$, $\mu_B$), which was not possible at lower energy heavy-ion accelerators.
Experimentally, that will be done by comparing the relative abundancies
between the different particles species to the ones expected by the
statistical models at equilibrium, which in turn furnishes the
parameters $T$ and $\mu_B$ that best describe the data.
These models have shown a good agreement with the data, for
example at RHIC~\cite{Andronic:2005yp}, which suggests that
chemical equilibrium is reached.
An inclusive measurement of all the particle species produced from
LHC \mbox{Pb-Pb} collisions will give insight on the question of chemical
equilibrium at an energy where hard processes are dominant.
It will be interesting to compare them
with the non-equilibrium scenario considered in
ref~\cite{Rafelski:2008an} which involves substantial strangeness
oversaturation. 
A comparison to small ``canonical'' systems (\mbox{p-p}) will also be very
important to detect volume effects in particle production.
The measurement of unflavoured and especially strange species will strongly constrain these models.

\begin{figure}[h]
\begin{center}
\includegraphics[scale=0.7]{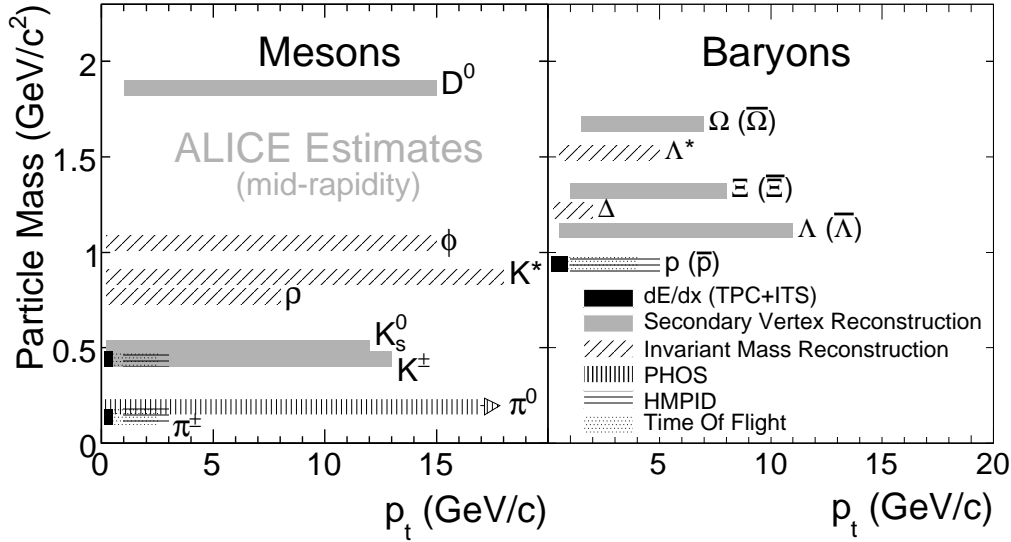}
\end{center}
\caption{\label{fig:ALICEpid}Transverse momentum range of identification of
mesons and baryons at mid-rapidity, expected for $10^7$ \mbox{Pb-Pb} central
events at the energy $\sqrt{s_{NN}}=5.5~{\rm TeV}$. (Note: these
ranges are sensitive to the models used in the simulations.)}
\end{figure}

Fig.~\ref{fig:ALICEpid} shows the range of transverse
momentum that ALICE will cover with the expected statistics of the first year of \mbox{Pb-Pb} data
taking, corresponding to $10^7$ central events.
At mid-rapidity, 
identification of charged particles is performed via their energy loss (ITS, TPC) and time
of flight (TOF) up to momenta $\sim 2~{\rm GeV}/c$ in a full azimuthal
coverage. The TRD will in addition separate electrons from pions.
At higher momentum the HMPID is used, extending the identification
range of kaons and protons up to 5~GeV$/c$ in a limited azimuthal coverage.
In addition, identification of particle decays is done via invariant mass analysis (for
resonances) plus topological methods for strange and charm secondary
vertices. These methods improve substantially the accessible $p_T$ range, up to more than 10~GeV$/c$.

\section{Expansion and hadronization}
As the fireball expands and cools
down, the quarks hadronize and the system freezes-out. What happens
during this expansion and hadronization phase can be inferred from the study of the
dynamical properties of the resulting hadrons.
In this respect, the study of elliptic flow brings interesting
information (see~\cite{Adams:2005dq}.) 
The measurement of the reaction plane is one of the
major issues in elliptic flow analysis. While the so-called ``event
plane'' method is sensitive to ``non-flow'' effects, the
``cumulant'' method doesn't give access to the reaction plane~\cite{Poskanzer:1998yz,Borghini:2001vi}.
A new possibility, based on both Lee-Yang Zeroes and event-plane methods has been
developed for ALICE; it provides the $v_2$ factor with a precision of $\sim 1\%$ (statistic error only), getting rid
of non-flow effets and auto-correlations, and the reaction plane with
a resolution factor of about 0.8~\cite{Bilandzic:2008nx,Kolk:2009}.

A good knowledge of the elliptic flow can supply substantial
information on the hadronization mechanisms. RHIC data have shown a
very good scaling of the $v_2$ factor with respect to the number of hadron
constituent quarks (two or three) as a function of transverse mass.
The coalescence models have been shown to provide one of the most accurate
explanations of this behaviour, and that has been corroborated by
baryon-over-meson ratio vs $p_T$ measurements~\cite{Fries:2008hs}.
In this context, the identification of strange secondary vertices
($K^0_S$, $\Lambda$, $\Xi$ and $\Omega$) plays an important role:
because their decay length is of several centimeters, their charged
decay modes can be reconstructed with topological methods, in which
single particle PID is not mandatory, and thus have no limitation
other than statistics.
As a result, the accessible $p_T$ range should go from $0.5$ to
at least 8-10~GeV$/c$ in the first \mbox{Pb-Pb} run of ALICE (see fig.~\ref{fig:ALICEpid}.)
This allows one to test the coalescence models and
identify the $p_T$ region above which fragmentation dominates
coalescence in hadron production.
Light resonances ($\rho$, $K^*$, $\phi$...) will give complementary
information on coalescence thanks to their range of identification in
$p_T$, but will also help probe chiral symmetry restoration~\cite{Pisarski:1981mq}
and the time between chemical and kinetic freeze-outs~\cite{Torrieri:2001ue}. In that
respect, the study of resonant-to-non-resonant particle ratios as a
function of event multiplicity and $p_T$ will be interesting.
It is expected that resonances can be identified up to $15~{\rm
GeV}/c$ in $p_T$, and recent studies show that PID is not mandatory
either~\cite{Badala:2008ga}, be it in \mbox{p-p} or heavy-ion collisions.

The interferometry method applied to identical particles is also a
powerful tool to probe the space-time evolution of the emitting source in
heavy-ions collisions, which will help constrain the hydrodynamical models.
The source radii observed at RHIC differ significantly from the
model predictions, possibly due to an incomplete hydrodynamic evolution of the fireball.
That measurement will therefore be fundamental at LHC. ALICE offers
great opportunities to perform these femtoscopic analyses thanks to
its identification performance at low $p_T$ and the expected high
event multiplicities. In such conditions, source radii up to 15~fm should be measurable~\cite{Kuhn:2008zza}.

\section{Event-by-event physics}

Fluctations of thermodynamical quantities from event to event are fundamental
to study the QCD phase transition and the order of this transition.
As a matter of fact, thanks to the identified particle multiplicities expected
in ALICE, various observables relative to the soft sector 
will be analysed on an event-by-event basis, as mentioned in section~\ref{sec:eventcharact}.
For example, conserved quantities such as the net electrical charge
may provide information on the degrees of freedom of the initial state
of the collision, and thus directly on the phase transition;
fluctuations in temperature (extracted from particle $p_T$ spectra) should indicate the universality (or not) of the freeze-out
temperature.

Other interesting information will be obtained from the event-by-event
analysis program of ALICE.
Recent simulations using the Hijing particle generator show that
the relative error on reconstructed $K/\pi$ and $p/\pi$ ratios, event
by event, should not exceed a few percent; that will allow one to detect unexpected fluctuations due to QGP effects.
In addition, the balance functions, whose width is directly related
to the correlation range between particles, can probe the
hadronization time. A new feature possible in ALICE will be the study
of the balance function as function of $p_T$ for different rapidity gaps~\cite{Christakoglou:2009sq}.

\section{Conclusions}

The ALICE experiment at LHC has a wide program dedicated to the soft
phenomena. Both in \mbox{p-p} and \mbox{Pb-Pb}, the event multiplicity and particle pseudorapidity distributions will
be the first physics measurements ALICE will achieve. With the data
collected in just a few hours of \mbox{Pb-Pb} beams, many questions related to the medium
equilibrium, its equation of state, its expansion, and phase transition will be
assessed. The expected large event multiplicities will give the possibility
to address many observables on an event-by-event basis and thus 
allow a novel approach, hardly possible at lower-energy accelerators, to investigate the underlying physics.
Obviously, all the soft physics results will have to be combined with
results coming from hard probes (including heavy flavours and jets), in order to reach a broad vision of
all the phenomena involved at these energies.

\section*{References}

\bibliography{iopart-num}

\end{document}